\documentstyle[aps,multicol]{revtex}
\input{epsf}
\begin{document}

\draft

\title{Phase-space picture of resonance creation and avoided crossings}
\author{T. Timberlake and L.~E. Reichl}
\address{Center for Studies in Statistical Mechanics and Complex
Systems\\
The University of Texas at Austin\\
Austin, Texas 78712}
\date{October 1, 2000}
\maketitle

\begin{abstract}
Complex coordinate scaling (CCS) is used to calculate resonance eigenvalues
and eigenstates for a system consisting of an inverted Gaussian potential
and a monochromatic driving field.  Floquet eigenvalues and Husimi
distributions of resonance eigenfunctions are calculated using two
different versions of CCS. The number of resonance states in this system
increases as the strength of the driving field is increased, indicating
that this system might have increased stability against ionization when the
field strength is very high.  We find that the newly created resonance
states are scarred on unstable periodic orbits of the classical motion. 
The behavior of these periodic orbits as the field strength is increased
may explain why there are more resonance states at high field strengths
than at low field strengths.  Close examination of an avoided crossing
between resonance states shows that this type of avoided crossing does not
delocalize the resonance states, although it may lead to interesting
effects at certain field strengths.

\end{abstract}


\begin{multicols}{2}
\narrowtext

\section{Introduction}
\label{gintro}

The study of time-periodic quantum systems has attracted considerable
interest in recent years.  One of the primary motivating factors for this
interest is the development of ultra-high intensity lasers, which can
produce electric fields within atoms that rival those produced by the
atomic nucleus.  Experiments with these ultra-intense lasers have led to the
discovery of many interesting new phenomena, such as high harmonic generation
\cite{murnane}.  Simple one-dimensional models of the interaction between
intense lasers and atoms have been shown to reproduce, at least
qualitatively, many of these interesting phenomena \cite{eberly}.  These models
are especially interesting because their
classical versions display chaotic motion
\cite{Reichl}.  In addition to providing insight into recent
experiments, the study of these models can also provide insight into
quantum-classical correspondence.

One of the more interesting phenomena observed in these systems is the
stabilization of atoms in intense laser fields.  Stabilization is
characterized by a \emph{decrease} in the probability for an electron to
ionize as the laser intensity is \emph{increased}.  This effect was first
discovered in theoretical studies of the interaction between high-frequency
lasers and atoms \cite{pont}, but recent experiments have verified that
stabilization occurs in real atoms \cite{deboer,frey}.  Studies of the
underlying classical dynamics of these systems using one- and
two-dimensional models have shown that the classical motion can often
account for the increased stability of the atom at higher laser intensities
\cite{frey,chism2}.

The study of time-periodic quantum models is usually carried out within the
context of Floquet theory \cite{Shirley}.  Floquet eigenstates are
eigenstates of the one-period time evolution operator and are the natural
states for describing time-periodic systems.  In some cases the Floquet
states of the system can be localized on stable structures in the classical
phase space \cite{yoshida}, and this can lead to stabilization because
these Floquet states have very long lifetimes.  In this case, stabilization
would also be predicted by the classical dynamics.  However, there are
often significant differences between the classical and quantum dynamics of
chaotic systems.  One of the most striking examples of this is scarring,
where quantum eigenstates have higher probability to be found near the
locations of unstable periodic orbits in the classical phase space
\cite{heller}.  The scarring of Floquet states on
unstable periodic orbits might make it possible for a quantum system to
exhibit stabilization even when the corresponding classical dynamics is
unstable.  Some earlier studies indicate that stabilization can
be associated with states that are scarred on unstable or weakly
stable periodic orbits \cite{sanders}. 

In this paper we examine a time-periodic system with one space dimension
which shows signs of stabilization.  In this system the number of localized
Floquet states, or resonance states, increases as the intensity of the
driving field is increased.  In Sec.  \ref{gmodel} we present the model and
discuss the classical dynamics as well as prior studies of the quantum
dynamics.  In Sec.  \ref{cscale} we describe two different versions of
complex coordinate scaling and compare their predictions in this system. 
In Sec.
\ref{newres} we investigate the relationship between the resonance states
and the classical dynamics of the system.  We find that the resonance
states that are created as the driving field is increased are associated
with unstable periodic orbits in the classical dynamics.  We give an explanation,
based on this association with periodic orbits, for why the number
of resonance states increases as the laser field is increased.

In Sec.  \ref{resac} we carry out a detailed study of an avoided crossing
between two resonance states.  Avoided crossings in time-periodic quantum
systems lead to significant changes in the structure of the Floquet states
\cite{latka,timberlake}.  Overlapping avoided crossings can
even lead to delocalization of Floquet states \cite{timberlake}.  Since
stabilization depends upon the Floquet states remaining localized near
(stable or unstable) classical structures, avoided crossings may play an
important role in destroying stabilization.  Finally, in Sec.  \ref{gconc}
we summarize our findings.

\section{Driven Inverted Gaussian Model}
\label{gmodel}

\begin{figure}
\epsfxsize=3.2in
\epsfysize=5.3in
\epsffile{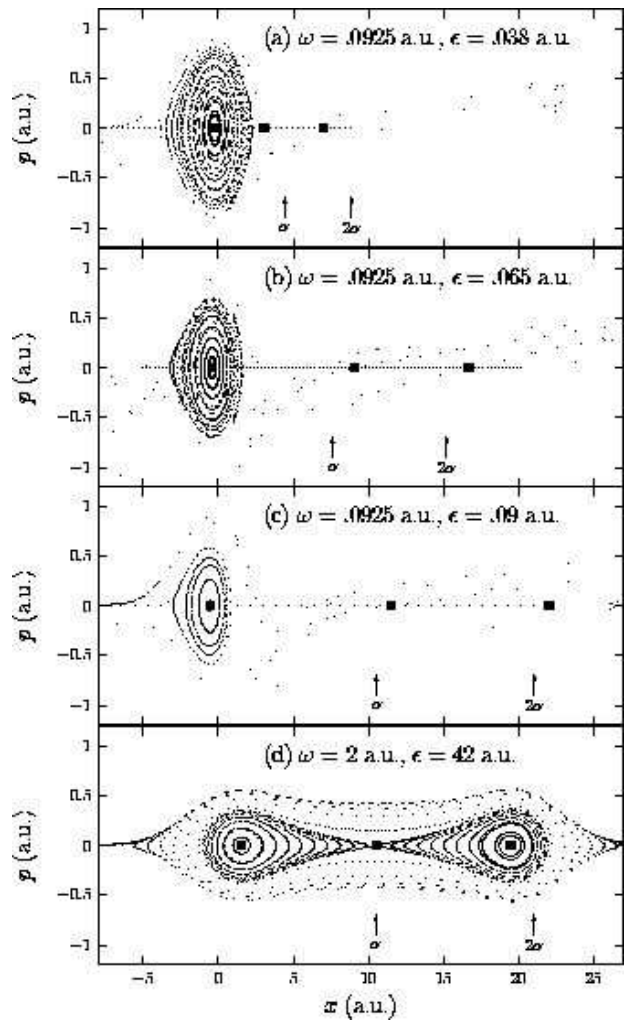}
\vglue 0.2cm
\caption{Strobe plots of the classical dynamics for the driven inverted
Gaussian system.  The initial 
conditions used to generate the plots all lie on the line $p=0$.  $\alpha$
is the classical excursion parameter for a free electron the field.  The 
locations of the periodic
orbits (stable and unstable) are indicated by filled squares.}
\label{igsp}
\end{figure}

The model we will study is an inverted Gaussian potential interacting with
a monochromatic driving field in the radiation gauge.  The Hamiltonian of the
system in atomic units (which are used throughout the paper) is
\begin{equation}
H=\frac{1}{2}\left(p-\frac{\epsilon}{\omega}\sin(\omega t)\right)^{2} - V_{0}
\exp(-(x/a)^{2})
\end{equation}
where $V_{0}=0.63$ a.u. and $a=2.65$ a.u.
It is useful to write this as
$H=H_{0}+V$, where
\begin{equation}
H_{0} = \frac{p^{2}}{2} - V_{0}\exp(-(x/a)^{2})
\end{equation}
and
\begin{equation}
V = -\frac{\epsilon}{\omega} p \sin(\omega t) + \frac{\epsilon^{2}}{2\omega^{2}}
\sin^{2}(\omega t).
\end{equation}
Figure \ref{igsp} illustrates the classical dynamics of this system for
driving frequency $\omega=.0925$ a.u.   The strobe plots in Fig. \ref{igsp}
are calculated by evolving a set of trajectories, all with initial momentum
$p=0$, over many cycles of the field and plotting the location of each
trajectory when $t=2\pi n/\omega$ (after each full cycle of the field). 
When no driving field is present the motion is regular and bounded
for negative energies.  Motion at positive energies
is unbounded.  Figs.  \ref{igsp}a, \ref{igsp}b, and 
\ref{igsp}c show the
classical strobe plots for $\epsilon=.038$, $.065$, and $.09$ a.u.,
respectively.  As $\epsilon$ is increased the region near $(x=0,p=0)$
remains stable, but the size of the stable region gets smaller as
$\epsilon$ is increased.  The filled squares in Fig.  \ref{igsp} indicate
the locations of the periodic points in the strobe plot and the arrows show
$x=\alpha$ and $x=2\alpha$, where $\alpha=\epsilon/\omega^{2}$ is the
excursion parameter of a free electron in the field.  At these parameter
values the periodic orbit at $(0,0)$ is stable while the other two periodic
orbits are unstable.  As $\epsilon$ is increased the unstable periodic
orbits move toward larger values of $x$.  For very high frequency driving
fields two of the periodic orbits can be stable while the third is
unstable.  This is illustrated in Fig.
\ref{igsp}d, which shows the classical strobe plot for $\omega=2$ a.u. and
$\epsilon=42$ a.u. The value of $\alpha$ in Fig.  \ref{igsp}d is the same
as in Fig.  \ref{igsp}c, but at the higher frequency the periodic orbit
located near $x=2\alpha$ is a stable elliptic orbit surrounded by regular
motion.  The periodic orbit at $x=\alpha$ is hyperbolic.

The quantum dynamics of this system has been the subject of several
investigations during the past decade.  The resonance states of this system
were first calculated by Bardsley and Comella in 1989 \cite{bardsley}. 
More recent studies have focused on high-harmonic generation (HHG) in this
system \cite{ben-tal2}.  It is the findings of Ben-Tal, Moiseyev,
and Kosloff \cite{ben-tal}, hereafter BMK, that have the most relevance to
our work.  They found that the number of resonance states in this system
increased as the field strength was increased over a certain range.  BMK
attempt to explain the creation of new resonance states as the field
strength is increased by analyzing the dynamics of the time-averaged system
in a reference frame that oscillates with a free electron in the driving
field, known as the Kramers-Henneberger or K-H frame
\cite{kramers}.  They found qualitative agreement in that the
number of bound states in the time-averaged potential increases as the
field strength is increased.  However, the quantitative agreement was not
very good.  This is not surprising since the time-averaged K-H description
is only accurate for very high frequency driving fields.  Since the
frequency of the driving field used by BMK and in this work ($\omega=.0925$
a.u.) is lower than the frequency of motion for two of the bound states in
the undriven system ($.4451$ a.u. and $.1400$ a.u.), the time-averaged K-H
description is not quantitatively accurate.  It is somewhat surprising that
the time-averaged K-H description is qualitatively accurate because the
classical motion of the system in the time-averaged K-H frame is stable
while the classical motion of the true system is largely unstable.  For
$\alpha > 1$ the time-averaged potential in the K-H frame is a double well
with minima separated by approximately $2\alpha$.  Motion in this double
well would be quite different from that seen in the strobe plots in Fig.
\ref{igsp}a-c (although it would closely resemble the motion shown in Fig. 
\ref{igsp}d, which is at a frequency that is high enough for the
time-averaged K-H description to be valid).  Our goal in this paper is to
find an alternative explanation for the creation of resonance states as
$\epsilon$ is increased in this system, an explanation that does not rely
on the time-averaged K-H description.

\section{Complex Coordinate Scaling}
\label{cscale}

In recent years the technique of complex coordinate scaling (CCS) has been
used extensively in the study of open quantum systems.  In this section we
will review two versions of complex coordinate scaling (standard and
exterior scaling) and show how these techniques can be used to compute the
resonance states of an open, time-periodic system.  Results from the
standard and exterior scaling versions are compared, for both
time-independent and time-dependent calculations.

\subsection{Standard complex coordinate scaling (CCS)}
\begin{figure}
\epsfxsize=3.2in
\epsfysize=3.7in
\epsffile{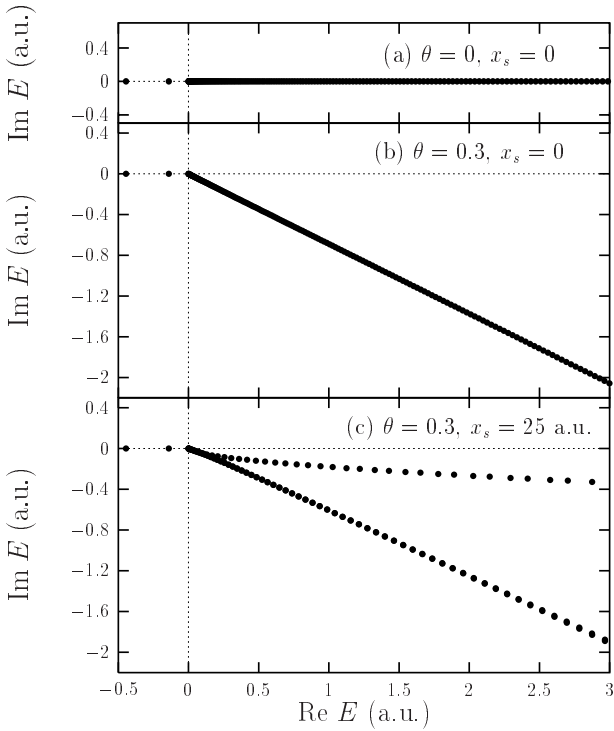}
\vglue 0.2cm
\caption{The complex-scaled energies of the undriven inverted Gaussian
system. The unscaled energies are shown in (a), the CCS energies in
(b), and the ECCS energies in (c).  The bound states of the system
have energies -0.4451, -0.1400, and -0.0001 a.u.  All calculations were performed
using a box size $L=200$ a.u. and 400 basis states.}
\label{gnrg}
\end{figure}

We first examine how the eigenvalues and eigenstates of a time-independent open system
can be calculated using standard complex coordinate scaling (CCS), a technique that
is examined in detail in
\cite{reinhardt,moiseyevres}.  In this paper we will use a basis of
particle-in-a-box states
for our calculations.  These states are defined by
\begin{equation}
\label{piab}
\langle x|n \rangle = \sqrt{\frac{2}{L}} \sin\left(\frac{n\pi x}{L}-\frac{n\pi}{2}\right),
\end{equation}
where $-L/2 \leq x \leq L/2$.  Calculations using CCS are performed just as they
are in traditional quantum mechanics, except that the coordinate is scaled
in the Hamiltonian so that $x \rightarrow xe^{i\theta}$ ($0 \leq \theta <
\pi/4$).  Scaling the coordinate in this fashion allows us to represent
resonance states, which are not in the Hilbert space, using square
integrable eigenfunctions.  As a result of this scaling the new
time-independent Hamiltonian is
\begin{equation}
\tilde{H_{0}}=H_{0}(xe^{i\theta}) = \frac{p^{2}e^{-2i\theta}}{2} - V_{0}\exp\left(-(xe^{i\theta}/a)\right).
\end{equation}
The kinetic energy operator is easily evaluated using the basis states in
Eq.  \ref{piab}. As long as our box is sufficiently large ($L >>
2x_{0}/\sqrt{cos(2\theta)}$) we find that
\begin{equation}
\label{vmatrix}
\langle m|-V_{0}\exp\left(-(xe^{i\theta}/a)^{2}\right)|n \rangle = V(m+n)-V(|m-n|)
\end{equation}
where
\begin{equation}
\label{vofj}
V(j) = \frac{V_{0}a\sqrt{\pi} e^{-i\theta}}{L} \exp\left(-\frac{j^{2}\pi^{2}
a^{2}e^{-2i\theta}}{4L^{2}}\right) \cos\left(\frac{j\pi}{2}\right).
\end{equation}
Once these matrix elements are calculated the $\tilde{H_{0}}$ matrix can be constructed.
Diagonalizing $\tilde{H_{0}}$ yields the energy eigenvalues of the time-independent system
as well as the eigenvectors
\begin{equation}
\label{nrgev}
|\psi_{i} \rangle = \sum_{n=1}^{N} c_{ni} |n\rangle.
\end{equation}
Fig.  \ref{gnrg}a shows the energy eigenvalues of $H_{0}$ calculated
without complex scaling ($\theta=0$).  The potential supports three bound
states at $E=-0.4451$, $-0.1400$, and $-0.00014$ a.u.  Without complex
scaling all eigenvalues lie on the real axis.  When the coordinate is
scaled, though, the Hamiltonian becomes non-Hermitian and it is possible
for eigenstates of the scaled system to have complex eigenvalues.  This can
be seen in Fig.  \ref{gnrg}b, which shows the eigenvalues calculated using
CCS with $\theta=0.3$.  The bound state eigenvalues remain on
the real axis but the positive-energy continuum states are rotated into the
lower half plane by an angle of $2\theta$.  It is this rotation of the
continuum that will allow us to identify resonances.  No resonances exist
for the system with Hamiltonian $\tilde{H_{0}}$, only bound states and a continuum of 
energy eigenvalues.

\subsection{Exterior complex coordinate scaling}
\label{extscale}
The basic idea of exterior complex coordinate scaling (ECCS) is to scale the
$x$ coordinate by a factor $e^{i\theta}$ as in CCS,
but only in the region $|x| \geq x_{s}$ where the
potential is zero.  Discontinuities at $\pm x_{s}$ are avoided by using
a smooth scaling relation $x \rightarrow F(x)$, where
\begin{equation}
F(x) = x + (e^{i\theta}-1)\left[x + \frac{1}{2\lambda} \ln\left( \frac{\cosh(\lambda
(x-x_{s}))}{\cosh(\lambda(x+x_{s}))} \right) \right]
\end{equation}
with $\lambda=5$ a.u. and $x_{s}=25$ a.u. This exterior scaling method is
given a thorough presentation in \cite{moiseyevres,moiseyevres2}.  Because
the potential is zero in the region where the coordinate is scaled, the
potential matrix elements can be calculated without any complex scaling
(i.e. using Eq.
\ref{vmatrix} and \ref{vofj}, but with $\theta=0$).  The scaled time-independent
Hamiltonian then becomes $\tilde{H_{0}} = H_{0} + V_{CAP}$, where
\begin{equation}
\label{vcap}
V_{CAP}(x) = V_{0}(x) + V_{1}(x)\frac{\partial}{\partial x} + V_{2}(x)\frac{\partial^{2}}
{\partial x^{2}}
\end{equation}
acts as a complex absorbing potential.  The coordinate-dependent factors
in Eq. \ref{vcap} are defined by
\begin{equation}
V_{0}(x) = \frac{1}{4} f^{-3}(x) \frac{\partial^{2} f}{\partial x^{2}} 
- \frac{5}{8} f^{-4}(x) \left(\frac{\partial f}{\partial x}\right)^{2},
\end{equation}
\begin{equation}
V_{1}(x) = f^{-3}(x) \frac{\partial f}{\partial x},
\end{equation}
and
\begin{equation}
V_{2}(x) = \frac{1}{2} (1-f^{-2}(x)),
\end{equation}
where $f(x) = \partial F / \partial x$.  Plots of $F(x)$, $V_{0}(x)$,
$V_{1}(x)$, and $V_{2}(x)$ are shown in
\cite{moiseyevres,moiseyevres2}. 

\begin{figure}
\epsfxsize=3.2in
\epsfysize=4.0in
\epsffile{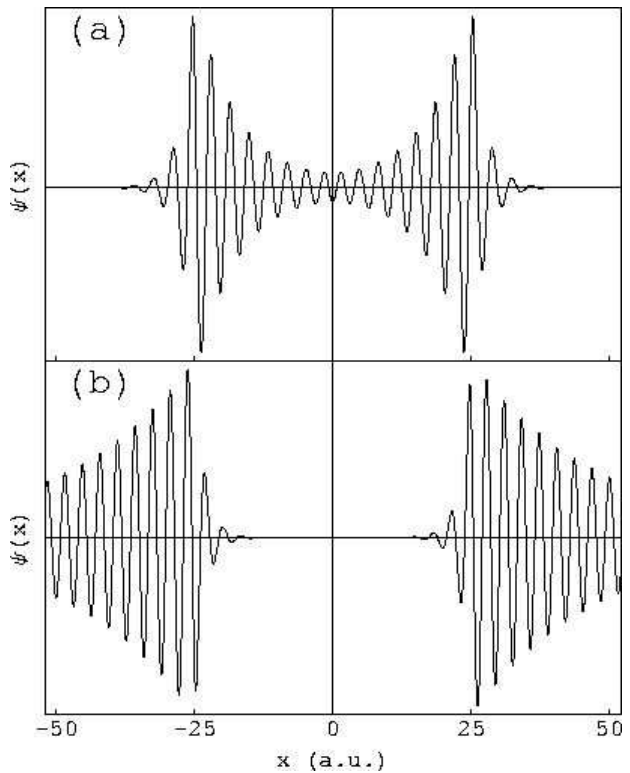}
\vglue 0.2cm
\caption{Wavefunctions of two ECCS continuum states.  The state shown
in (a) is a partially-scaled state whose eigenvalue is rotated by less than 
$2\theta$ from the real axis.  The state in (b) is a fully-scaled state whose 
eigenvalue is rotated the full $2\theta$ from the real axis.  The partially-
scaled state is localized between $-x_{s}$ and $x_{s}$ ($x_{s}=25$ a.u.), while
the fully-scaled state is almost excluded from this region.  Both states are peaked
near $\pm x_{s}$.}
\label{cwf}
\end{figure} 

We will again use a basis of particle-in-a-box
states to calculate $\tilde{H_{0}}$.  Matrix elements for the
potential energy term are calculated without any complex scaling, while
the kinetic energy and $V_{CAP}$ matrix elements are calculated numerically.  Diagonalizing
$\tilde{H_{0}}$ gives the complex energy eigenvalues for the exterior
scaled system, which are shown in Fig.  \ref{gnrg}c.  Note that the bound
state eigenvalues are still on the real axis and most of the continuum
states have been rotated into the lower half plane by $2\theta$.  However,
several of the positive energy states have been rotated into the lower half
plane by considerably less than $2\theta$.  We refer to these as ``partially
scaled'' continuum states.  Fig.  \ref{cwf} shows the wavefunction of one
fully scaled continuum state and one partially scaled continuum state.  The
partially scaled state is strongly peaked near $x=x_{s}$ and it is non-zero
only within the region $-x_{s} \leq x \leq x_{s}$, while the fully scaled
state is zero within this region.  As $x_{s}$ is decreased toward $0$, the
number of partially scaled states decreases.  At $x_{s}=0$ the ECCS
eigenvalues exactly match the CCS eigenvalues as expected.

Since we will want to examine the structure of the resonance wavefunctions
in the periodically driven system it is important first to examine the structure
of the eigenstates of $\tilde{H_{0}}$.  We will examine
the structure of the eigenstates of $\tilde{H_{0}}$ by calculating Husimi distributions
\cite{Husimi} for each of the three bound states.  An Husimi
distribution is a quasi-probability distribution of a quantum state in the phase space.  The
Husimi distribution (HD) of a quantum wavefunction $\Psi(x)$ is defined as
\begin{equation}
\label{huseq}
G(x_{0},p_{0}) = \left| \left(\frac{1}{\pi\sigma^{2}}\right)^{1/4} 
\int_{-\infty}^{-\infty} e^{-(x-x_{0})^{2}/2\sigma^{2} - ip_{0}x} \Psi(x) dx \right|^{2}.
\end{equation}

\begin{figure}
\epsfxsize=3.2in
\epsfysize=3.1in
\epsffile{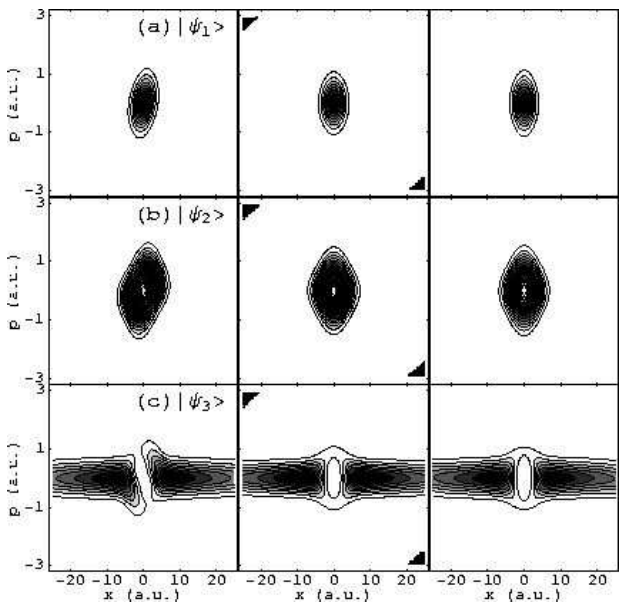}
\vglue 0.2cm
\caption{Husimi distributions of the bound states of the inverted
Gaussian system.  The distributions in the left column are of the CCS bound
states.  The distributions in the center column are CCS bound states
rotated back into the real coordinate frame.  The right column shows the
distributions for the ECCS ($x_{s}=25$ a.u.) bound states.  The scaling
angle is $\theta=0.3$ for all plots.}
\label{hust}
\end{figure}

Husimi distributions for the three bound states of $\tilde{H_{0}}$ are
shown in Fig.  \ref{hust}.  The left-hand column shows the HDs of the
eigenstates, $\psi_{i}(x)$, of $\tilde{H_{0}}$ obtained using
CCS. The center column shows the HDs for the eigenstates
$\psi_{i}(xe^{i\theta})$, which is $\psi_{i}(x)$ rotated back to the real
coordinate frame \cite{voro}.  The right-hand column shows the HDs of the eigenstates
that were calculated using ECCS. Note that the center and right-hand
columns agree quite well, while the left-hand column shows significant
differences.  The Husimi distributions in the right-hand column match those
that are found without complex scaling the Hamiltonian and hence they are
the correct distributions.  The distributions in the center column are
correct except for the patches of black in the top-left and bottom-right
corners.  These patches show where the Husimi distribution blows up because
the wavefunction is not square integrable. 

\subsection{Floquet calculations}

As we have seen, complex coordinate scaling can be used to calculate the
first $N$ energy eigenstates of $\tilde{H_{0}}$.  These eigenstates can
then be used as a basis to compute the one-period time evolution (Floquet)
matrix, $\hat{U}(T)$, for the driven system.  This matrix is calculated by
numerically integrating the time-dependent Schr\"{o}dinger equation $N$
times from $t=0$ to $t=T=2\pi/\omega$ with initial conditions
$|\Psi(t=0)\rangle = |\psi_{i}\rangle$, where $|\psi_{i}\rangle$ is the
$i$th energy eigenstate of $\tilde{H_{0}}$.  Diagonalization of this matrix
gives the Floquet eigenvalues and eigenstates (in the basis of
eigenstates of $\tilde{H_{0}}$).

The time-dependent Schr\"{o}dinger equation for the driven inverted Gaussian
system is
\begin{equation}
i\hbar \frac{\partial}{\partial t}|\Psi\rangle = \tilde{H_{0}}|\Psi\rangle - 
\frac{\epsilon}{\omega} \tilde{p}\sin(\omega t)|\Psi\rangle + 
\frac{\epsilon^{2}}{2\omega^{2}} \sin^{2}(\omega t) |\Psi\rangle
\end{equation}
where $\tilde{p}$ is the complex scaled momentum operator.  Since all
computations are performed in a basis of eigenstates of $\tilde{H_{0}}$ we
must first calculate the matrix elements of $\tilde{p}$.  In calculating
these matrix elements it is critical to recognize that $\tilde{H_{0}}$ is
not a Hermitian matrix and thus its eigenvectors do not have the usual
properties that eigenvectors of Hermitian matrices have.  One cannot obtain
the left eigenvectors of a non-Hermitian matrix simply by taking the
complex conjugate of the right eigenvectors.  In our case, $\tilde{H_{0}}$
is complex symmetric and the coefficients of the left eigenvectors are
equal to ({\it not} complex conjugates of) the coefficients of the right
eigenvectors, so
\begin{equation}
\langle \psi_{i}| = \sum_{n=1}^{N} c_{ni} \langle n |.
\end{equation}
The normalization of the eigenvectors is also different.  For our complex symmetric
matrix the eigenvectors should be normalized so that the sum of the squares of the
$c_{ni}$'s is $1$, rather than the sum of the absolute squares.  With this in mind
we can calculate the matrix elements for $\tilde{p}$ using
\begin{equation}
\langle\psi_{i}|\tilde{p}|\psi_{j}\rangle = \sum_{m=1}^{N} \sum_{n=1}^{N} c_{mi} c_{nj}
\langle m|\tilde{p}|n\rangle.
\end{equation}
The $\langle m|\tilde{p}|n\rangle$ are easy to calculate when CCS is used
and $\tilde{p}$ is simply $pe^{-i\theta}$.  However, when ECCS is used
those matrix elements are calculated numerically using
\begin{equation}
\langle m|\tilde{p}|n\rangle = \frac{-i\hbar\pi n}{L^{2}} \left[P(m+n)+P(m-n)\right]
\end{equation}
where
\begin{equation}
P(k) = \int_{-L/2}^{L/2} \sin\left( \frac{k\pi x}{L} - \frac{k\pi}{2}\right) f^{-1}(x) dx
\end{equation}
and $f(x)$ is defined in Sec. \ref{extscale}.

\begin{figure}
\epsfxsize=3.2in
\epsfysize=5.6in
\epsffile{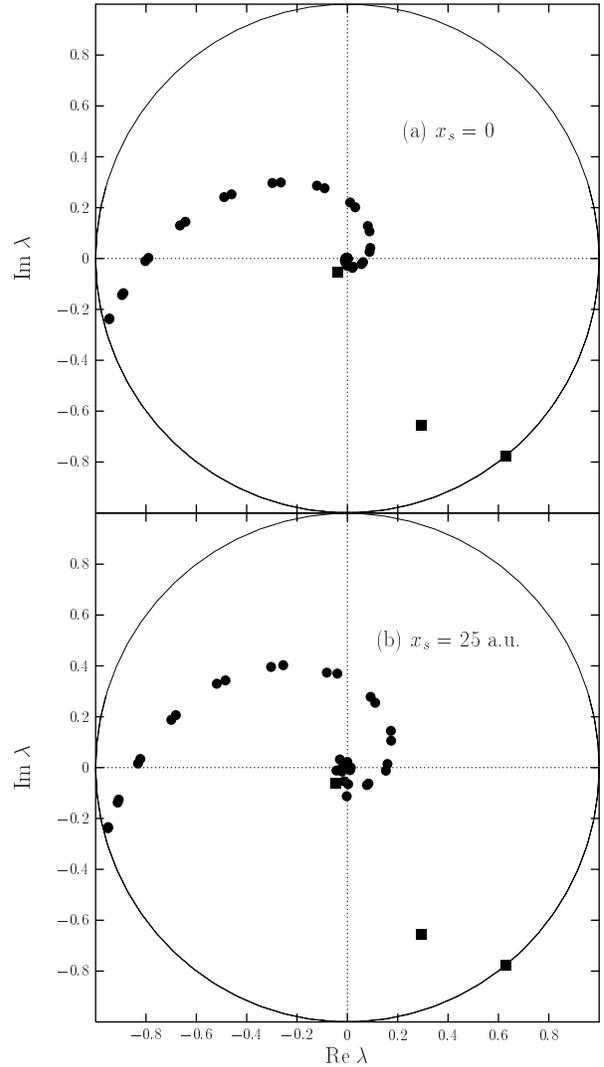}
\vglue 0.2cm
\caption{Floquet eigenvalues for the driven inverted Gaussian system with
$\omega=0.0925$ a.u. and $\epsilon=0.038$ a.u. The eigenvalues calculated
using CCS are shown in (a).  In (b) the ECCS
eigenvalues are shown.  The scaling angle is $\theta=0.3$ for both sets of
eigenvalues.  Resonance states are indicated by filled squares, while
continuum states are indicated by filled circles.}
\label{qev.038}
\end{figure}

By numerically integrating the Schr\"{o}dinger equation over one cycle of the driving
field we can construct the Floquet matrix as described above.  This matrix is then
numerically diagonalized to give the Floquet eigenvalues and eigenstates.  Since
the Floquet eigenstates are calculated in a basis of eigenstates of $\tilde{H_{0}}$
we can write them as
\begin{equation}
|q_{\beta}\rangle = \sum_{i=1}^{N} d_{i\beta}|\psi_{i}\rangle.
\end{equation}
Because they are eigenstates of the one-period time evolution operator (Floquet matrix)
we can write
\begin{equation}
\hat{U}(T)|q_{\beta}\rangle = e^{-iq_{\beta}T}|q_{\beta}\rangle
\end{equation}
where $q_{\beta}$ is the quasienergy of the state $|q_{\beta}\rangle$. 
Because the Hamiltonian $\tilde{H_{0}}$ is not Hermitian, the time
evolution operator is not unitary.  This means that the Floquet eigenvalues
do not necessarily have unit modulus, and thus the quasienergies
$q_{\beta}$ are in general complex.  We can write the quasienergies as $q_{\beta} =
\Omega_{\beta} + i\Gamma_{\beta}/2$, where $\tau_{\beta}=1/\Gamma_{\beta}$
is the lifetime of the state $|q_{\beta}\rangle$.  Resonance states are
easily identified by plotting the Floquet eigenvalues, which we will denote
as $\lambda_{\beta} = \exp(-iq_{\beta}T)$.  Fig.  \ref{qev.038} shows the
Floquet eigenvalues calculated using both CCS and ECCS for the driven
Gaussian system with $\omega=0.0925$ a.u. and $\epsilon=0.038$ a.u. The
eigenvalues that were found with the CCS method form a well-defined spiral
from the origin out to the edge of the unit circle.  These states are
indicated by filled circles in Fig.
\ref{qev.038}a.  Resonance states are indicated by filled squares and lie off of the
continuum spiral.  The continuum spiral is not as well-defined when the ECCS method
is used, as shown in Fig. \ref{qev.038}b.  However, only a few eigenvalues near the
origin appear to fall out of the spiral.  This could cause some difficulty in 
identifying broad (short-lived) resonances, but narrow (long-lived) resonances can
still be easily identified.  Fig. \ref{qev.038} shows that CCS and
ECCS appear to give the same resonance eigenvalues.

The resonance eigenvalues should be independent of the scaling angle
$\theta$, while the continuum eigenvalues rotate around the origin as
$\theta$ is changed.  However, when calculations are performed using a
finite basis the resonance eigenvalues will be weakly dependent upon
$\theta$ \cite{alon}.  To accurately determine the quasienergies
(and hence the lifetimes) of these states it is important to optimize
$\theta$ by finding the stationary point of each resonance eigenvalue as
$\theta$ is changed.  However, since our goal is not an accurate
quantitative determination of eigenvalues or lifetimes but a qualitative
understanding of the relationship between the quantum dynamics and the
classical motion, it is not critical that $\theta$ be optimized for our
calculations.  Optimizing $\theta$ presents a problem in this type of study
because the optimal value of $\theta$ is generally different for different
resonance states.  We wish to study all of the resonance states of the
system at the same time and it is impossible to optimize $\theta$ for all
resonance states within a single calculation of the Floquet matrix.  We
find that changing $\theta$ between $0.3$ and $0.7$ results in no visible
change in the plots of the resonance eigenvalues.  There is also no visible
change in the Husimi distribution of the states.  Not optimizing $\theta$
may lead to slight inaccuracies in the calculated lifetimes for the
resonance states, but we find that the error in the lifetimes is no greater
than $\pm 0.1T$ which is acceptable for our purpose here.

\begin{figure}
\epsfxsize=3.2in
\epsfysize=3.1in
\epsffile{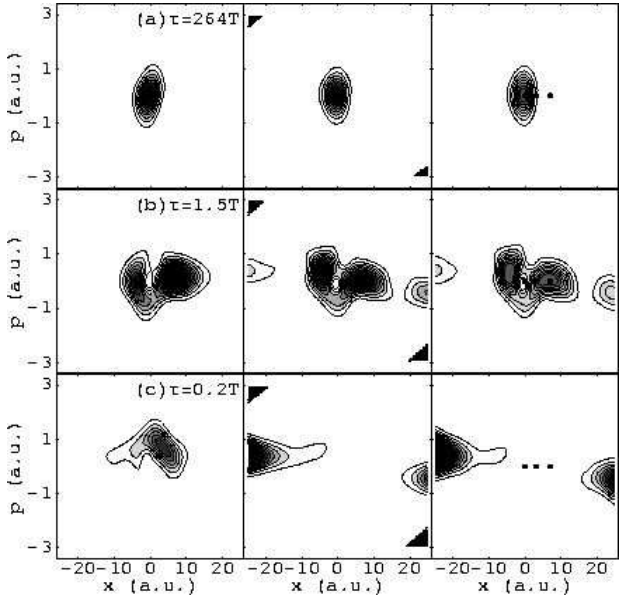}
\vglue 0.2cm
\caption{Husimi distributions of the resonance states at
$\epsilon=0.038$ a.u. The left column shows the Husimi distributions of the
states computed using CCS. The center column shows the distributions of the
CCS states rotated back into the real coordinate frame.  The right column
shows the distributions of the ECCS states ($x_{s}=25$ a.u.).  In the
right-hand column the locations of the periodic orbits are indicated by
filled circles and the contour plots have been lightened to allow the
circles to show.  The scaling angle is $\theta=0.3$ for all plots.  The
lifetimes for each state are given in units of the driving period
$T=2\pi/\omega$.}
\label{hus.038}
\end{figure}

In Figure \ref{hus.038} we show the Husimi distributions of the three
resonance states indicated in Fig.  \ref{qev.038}.  As in Fig.  \ref{hust}
the left-hand column shows CCS states, the middle column shows the rotated
CCS states, and the right-hand column shows ECCS states.  Again we find
agreement between the middle and right-hand columns, while the left-hand
column is inaccurate (as expected).  Lifetimes of the three states are
indicated in units of the driving period $T=2\pi/\omega$.  Filled circles indicate
the locations of the classical periodic orbits.  The resonance
state with the longest lifetime is almost indistinguishable from the ground
state of the undriven system shown in Fig.  \ref{hust}a.  The state shown
in Fig.  \ref{hus.038}b has a much shorter lifetime and is beginning to
elongate toward the positions of the unstable periodic orbits, with a peak of 
probability near the periodic orbit at ($x=6.98$ a.u., $p=0$).  The state
shown in Fig.  \ref{hus.038}c has a very short lifetime and its Husimi
distribution is similar to that of a continuum state.

\begin{figure}
\epsfxsize=3.2in
\epsfysize=5.6in
\epsffile{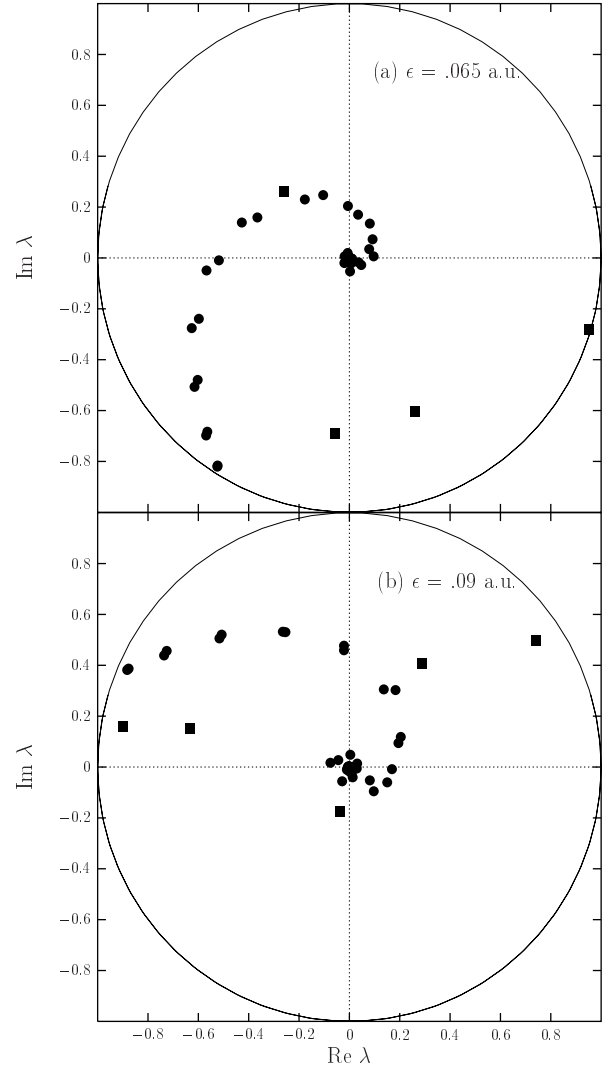}
\vglue 0.2cm
\caption{Floquet eigenvalues for $\omega=0.0925$ a.u. and two different field
strengths.  The eigenvalues are
calculated using ECCS with $\theta=0.3$ and $x_{s}=25$ a.u.
Resonance states are indicated by filled squares and continuum states by
filled circles.  At these higher field strengths the number of resonance states
is greater than at $\epsilon=.038$ a.u.}
\label{qevh}
\end{figure}

\section{Resonance creation and scarring}
\label{newres}

Because the rotated CCS states blow up for large $|x|$ we choose to perform
the remainder of our calculations using ECCS only.  As long as our unscaled
region ($-x_{s}< x < x_{s}$) is large enough to contain the phase
space region in which we are interested, we will not have to worry about
any inaccuracy associated with the scaling procedures.  We choose
$x_{s}=25$ a.u. which is large enough to include the locations of the
periodic orbits at all of the field strengths we study.

Figure \ref{qevh} shows the Floquet eigenvalues for $\epsilon=0.065$ and
$0.09$ a.u. Again, resonance states are indicated by filled squares while
continuum states are indicated by filled circles.  Comparing these plots
with Fig.  \ref{qev.038} we see that the number of resonance states
increases as $\epsilon$ is increased, from only three at $\epsilon=0.038$
a.u. to five at $\epsilon=0.09$ a.u. This is in agreement with BMK
\cite{ben-tal}.  In the classical system, however, the stable structure
near $(x=0,p=0)$ gets smaller as $\epsilon$ is increased.  If the resonance
states were associated with this stable classical structure then some of
the resonances should disappear as $\epsilon$ is increased.  Instead, the
opposite behavior is found.  To find the explanation for the increase in
the number of resonance states we examine the Husimi distributions of the
resonance states indicated in Fig.  \ref{qevh}.

\begin{figure}
\epsfxsize=3.2in
\epsfysize=3.1in
\epsffile{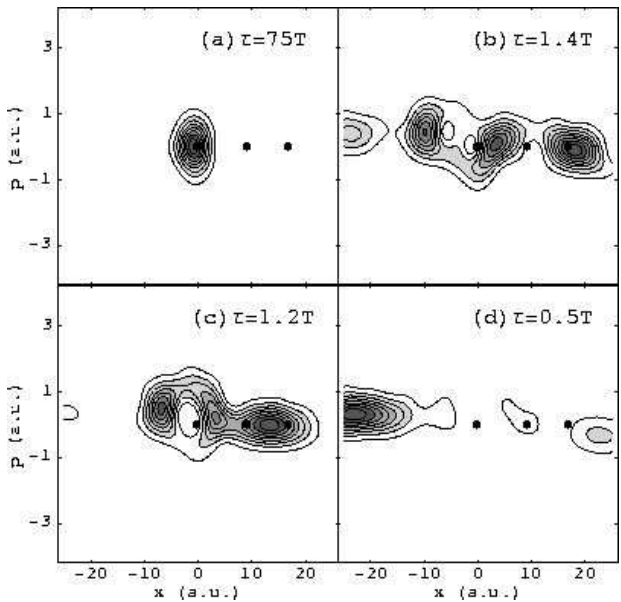}
\vglue 0.2cm
\caption{Husimi distributions for the four resonance states 
at $\epsilon=0.065$ a.u.  Lifetimes for each state
are given in units of the driving period $T=2\pi/\omega$.  The locations of
the periodic orbits are indicated by filled circles.}
\label{hus.065}
\end{figure}

Figure \ref{hus.065} shows HDs for the four resonance states at
$\epsilon=0.065$ a.u. Lifetimes for the states are indicated in units of
the driving period and the positions of the periodic orbits are indicated
by filled circles.  The state with the longest lifetime looks very much
like the ground state of the undriven system.  The states shown in Figs. 
\ref{hus.065}b and \ref{hus.065}c show some similarities to the excited
states of the undriven system, but they have both been elongated in the
direction of the unstable periodic orbits.  The state shown in Fig. 
\ref{hus.065}b has a probability peak near the unstable periodic orbit at
($x=16.67$ a.u., $p=0$), while the state shown in Fig.  \ref{hus.065}c has
a peak between the two unstable orbits.  These two states appear to have
become at least partially associated with the unstable periodic orbits. 
The state shown in Figure \ref{hus.065}d is the newly created resonance and
it has the shortest lifetime of the four.  It has a modest peak near the
periodic orbit at ($x=9.09$ a.u., $p=0$).

\begin{figure}
\epsfxsize=3.2in
\epsfysize=3.1in
\epsffile{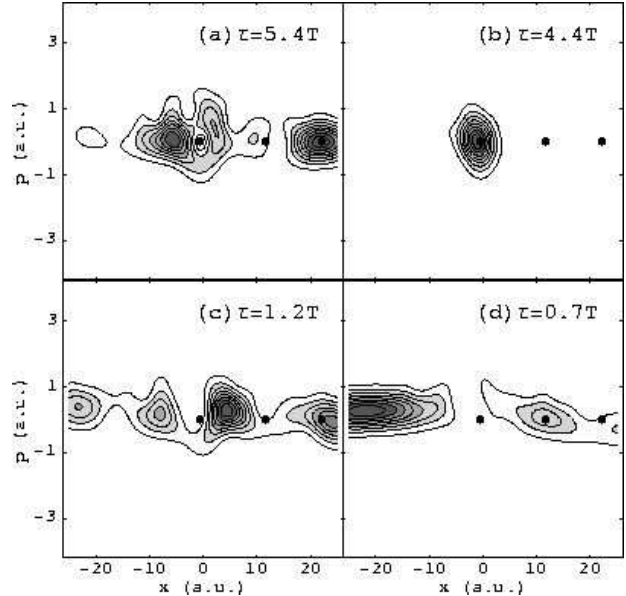}
\vglue 0.2cm
\caption{Husimi distributions for the four of the five resonance
states at $\epsilon=0.09$ a.u.  The resonance
state with the shortest lifetime is not shown because its Husimi
distribution is indistiguishable from that of a continuum state.  Lifetimes
for each state are given in units of the driving period $T=2\pi/\omega$. 
The locations of the periodic orbits are indicated by filled circles.}
\label{hus.09}
\end{figure}

Figure \ref{hus.09} shows HDs for four of the five resonance states at
$\epsilon=0.09$ a.u. The state that closely resembles the undriven ground
state (Fig.  \ref{hus.09}b) no longer has the longest lifetime.  Instead,
the longest-lived state resembles the first excited bound state of $H_{0}$,
but with additional peaks near the periodic orbits at $x=11.52$ a.u. and
$x=22.04$ a.u. The state shown in Fig.  \ref{hus.09}d is similar to the
state shown in Fig.  \ref{hus.065}d, but with a more prominent peak near
the periodic orbit at $x=11.52$ a.u. Note that the lifetime of this state
is also greater than that of the state shown in Fig.  \ref{hus.065}d.

At low values of $\epsilon$ all of the resonances have their probability
concentrated near $(x=0,p=0)$.  At these low values of $\epsilon$ the two
unstable periodic orbits are located close to the stable orbit near
$(x=0,p=0)$.  If any resonance state was associated with the unstable
periodic orbits at such low field strengths it would be difficult to tell
from its Husimi distribution.  As $\epsilon$ is increased the unstable
periodic orbits move toward larger values of $x$ and some of the resonance
states begin to spread in that direction as well.  At moderate values of
$\epsilon$ some states show peaks near the periodic orbit that is farthest
from $(x=0,p=0)$, close to $x=2\alpha$.  Only at high values of $\epsilon$
do we begin to see a state that is peaked on the unstable orbit that is
closest to $(x=0,p=0)$, near $x=\alpha$.  We believe that it is the
association between the resonances and the unstable periodic orbits that
explains the creation of resonance states as $\epsilon$ is increased.  At
low $\epsilon$ all three periodic orbits are too close together to support
many quantum states because they all occupy essentially the same region of
phase space.  As $\epsilon$ is increased the unstable periodic orbits move
away from the stable one and from each other.  This allows quantum states
to be associated with these unstable orbits without occupying the same
region of phase space as the states associated with the stable orbit, so
new resonance states are created.  It is the scarring of resonance states
on unstable periodic orbits of the classical system that accounts for the
increase in the number of resonance eigenstates, even as the stable region
in the classical phase space is diminished.

The behavior we see here would be unlikely to stabilize the ground state of the 
undriven system against ionization in a high intensity field.  This is because
the lifetime of the resonance state that seems most closely related to the ground state (shown in
Figs. \ref{hus.038}a, \ref{hus.065}a, and \ref{hus.09}b) decreases as $\epsilon$
is increased.  However, the observed behavior could lead to stabilization for an
excited state of the undriven system.  The excited states have most of their probability
away from $(x=0,p=0)$ and would thus overlap with resonance states
that are not peaked at that point.  Since these resonances grow in number and increase
their lifetimes as $\epsilon$ is increased, an excited state of the undriven system
may become stabilized against ionization as $\epsilon$ is increased.

\section{Avoided crossings between resonances}
\label{resac}

Avoided crossings between resonance eigenvalues have been identified in this system
\cite{ben-tal2,ben-tal}.  Avoided crossings between Floquet eigenvalues play
an important role in multi-photon ionization \cite{gontier} and the delocalization
of Floquet eigenstates \cite{timberlake}.  In this section we investigate the quantum
dynamics at one avoided crossing to determine if it leads to delocalization of the
resonance eigenstates.  Delocalization is closely related  to ionization in
these systems because long-lived resonance states can only exist if they are localized
within the interaction region.  If avoided crossings lead to delocalization they would
also lead to a decrease in the lifetime, and eventually the destruction, of the resonance
states.  

\begin{figure}
\epsfxsize=3.2in
\epsfysize=2.9in
\epsffile{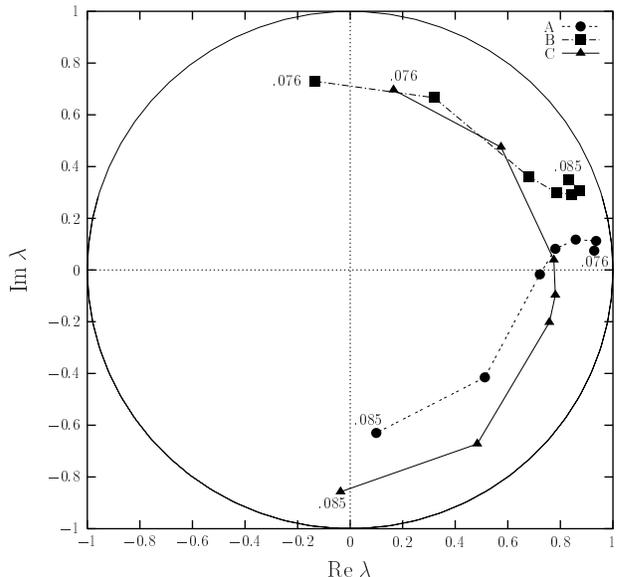}
\vglue 0.2cm
\caption{Floquet eigenvalues of three resonance states (labeled $A$, $B$, and $C$) that are
involved in an avoided crossing around $\epsilon=0.0805$ a.u. The points
show the eigenvalues for $\epsilon=.076, .078, .08, .0805, .081, .083,$ and
$.085$ a.u. The numbers shown in the plot indicate the field strengths at
the end points of each eigenvalue sequence.  The lines are intended only as
an aid to the eye.}
\label{qevcross}
\end{figure}

Figure \ref{qevcross} shows the Floquet eigenvalues of three resonance
states at several field strengths between $\epsilon=0.076$ and $0.085$
a.u. Two of these states (labeled $A$ and $B$ in Fig.  \ref{qevcross} and indicated
by filled circles and squares, respectively) are
involved in a prominent avoided crossing at a field strength of about
$\epsilon=0.0805$ a.u. The third resonance eigenvalue (labeled $C$ and indicated
by filled triangles) passes close by the
other two at this field strength, but it is not clear from Fig. 
\ref{qevcross} if that state is involved in the avoided crossing.  

\begin{figure}
\epsfxsize=3.2in
\epsfysize=3.1in
\epsffile{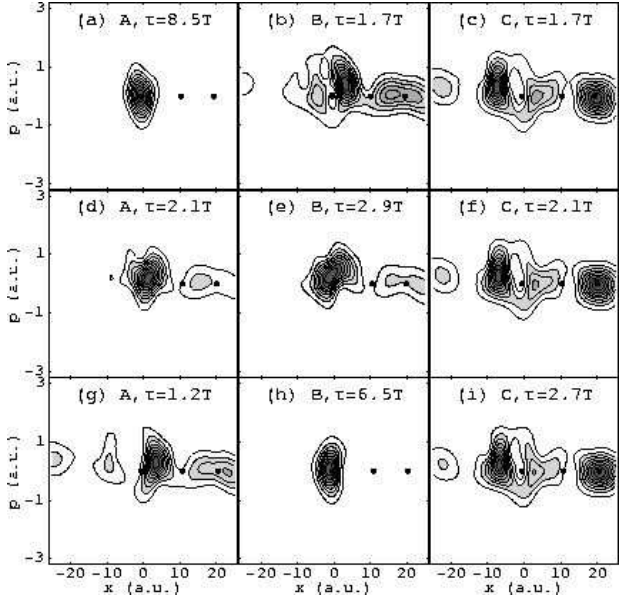}
\vglue 0.2cm
\caption{Husimi distributions for the states involved in the
avoided crossings shown in Fig. \ref{qevcross}.  The labels $A$, $B$, and $C$
correspond to the labels in Fig. \ref{qevcross}.  Lifetimes for each state are
given in units of the driving period $T=2\pi/\omega$.  The top row (a-c) shows Husimi
distributions for the three resonance states at $\epsilon=.078$ a.u.  The second
row (d-f) shows the distributions at $\epsilon=.0805$ a.u.  The bottom row (g-i)
shows the distributions at $\epsilon=.083$ a.u.  States $A$ and $B$ appear to 
exchange their structure as they pass through the avoided crossing.  State $C$
does not appear to undergo any major changes in its structure, but its lifetime
increases dramatically as it passes through the avoided crossing. The locations of
the periodic orbits at each field strength are indicated by filled circles.}
\label{huscross}
\end{figure}

To determine the effect of this avoided crossing on the resonance states we
examine the Husimi distributions of the states $A$, $B$, and $C$ shown in
Figure \ref{huscross}.  As $\epsilon$ is increased from $0.078$ a.u. to
$0.0805$ a.u. the states $A$ and $B$ undergo strong mixing with each other. 
When the field strength is increased to $0.083$ a.u. we find that states
$A$ and $B$ have completely exchanged their structure.  State $C$ does not
appear to have any significant structural changes in this range of field
strengths.  However, it should be noted that state $C$ has a significant
increase in its lifetime as $\epsilon$ is increased from $0.078$ a.u. to
$0.083$ a.u. States $A$ and $B$ exchange lifetimes as well as structure,
but the lifetimes of both states at $\epsilon=0.083$ a.u. are somewhat
smaller than the corresponding lifetimes at $\epsilon=0.078$ a.u. It may be
that state $C$ somehow gains stability at the expense of states $A$ and
$B$, even though it does not appear to pick up any of the structure of
those states.  Note that there are slight differences between the Husimi 
distributions of states $A$ and $B$ at $\epsilon=0.073$ a.u. and the 
corresponding distributions at $\epsilon=0.083$ a.u., but these may be due to
a small amount of mixing with state $C$ or with continuum states.  

Figure \ref{huscross} does not reveal any significant increase in the
delocalization of any of the resonance states.  It does indicate that an
avoided crossing of this type might lead to changes in the lifetimes of the
resonance states, but although two states have their lifetimes decreased
the third has its lifetime increased.  There is no clear indication that
this type of avoided crossing contributes to the destruction of the
resonance states that might prevent stabilization at very high intensities
of the driving field.  We expect that the destruction of resonance states
occurs primarily as the result of coupling between a resonance state and
the continuum rather than between resonance states \cite{timber3}. 

Avoided crossings between resonance states may play an important role in
other phenomena in this system.  For example, it has been shown in bound
systems that avoided crossings between Floquet eigenstates can result in
increased high harmonic generation (HHG).  Avoided crossings contribute to
increased HHG in two ways.  During the turn-on of a laser field avoided
crossings can put the quantum system into a superposition of Floquet states
that may emit radiation at higher frequencies than would be emitted by a
single Floquet state \cite{chism}.  Avoided crossings also contribute to
HHG by spreading the Floquet states over a wider range of energy, thus
allowing a single Floquet state to emit higher frequency radiation.  For
the type of avoided crossing observed here the states are only delocalized
near the exact field strength at which the avoided crossing occurs, because
at this field strength the Floquet states have mixed their structure
\cite{timberlake}.  At that particular field strength,
though, this effect could lead to increased HHG. In fact, increased HHG
has been observed in previous studies of the avoided crossings in this
system
\cite{ben-tal2}.

\section{Conclusion}
\label{gconc}

We utilize the complex coordinate Floquet method (CCFM) to calculate
resonances for an inverted Gaussian potential driven by a monochromatic
field.  As has been previously observed, we find that the number of
resonance states increases as the field strength is increased.  This
behavior in the quantum system seems to be opposite to what is observed in
the classical system, where the dynamics becomes increasingly unstable as
the field strength is increased.  An examination of the Husimi
distributions of the resonance states in this system shows that the newly
created resonances states are associated with unstable periodic orbits in
the classical motion.  This scarring of the eigenstates on unstable
periodic orbits has been seen in other systems and it represents one of the
most significant deviations of quantum dynamics from the corresponding
classical dynamics.  In this system, resonance eigenstates are scarred on
unstable periodic orbits and the movement of the periodic orbits in the
phase space allows for the creation of new resonance states as the field
strength is increased.  The creation of these new resonance states might
help to stabilize the system against ionization in intense fields.

An avoided crossing between resonance states is closely examined and it is
found that, although two states do exchange their structure during the
avoided crossing, there is no significant delocalization of the resonance
states as a result of this avoided crossing.  Coupling to the continuum is
more likely to destroy resonance states than is coupling between two
resonance states.  Avoided crossings of the type studied here, although
they may not lead to faster ionization, could lead to many interesting
effects such as increased high harmonic generation.

\acknowledgments

The authors wish to thank the Welch Foundation Grant No. F-1051 and
DOE Contract No. DE-FG03-94ER14465 for partial support of this work.
We also thank the University of Texas at Austin High
Performance
Computing Center for their help and the use of their computer facilities.

\appendix

\end{multicols}

\end{document}